# Devitrification of a glass-like arrested ferromagnetic phase in $La_{0.5}Ca_{0.5}MnO_3$.

A. Banerjee, Kranti Kumar and P. Chaddah.

*UGC-DAE Consortium for Scientific Research, University Campus, Khandwa Road, Indore-452017, INDIA.*

## Abstract

Magnetization measurements in $La_{0.5}Ca_{0.5}MnO_3$ manganite show that the high-temperature long-range ferromagnetic-metallic phase transforms to antiferromagnetic-insulating phase, although a fraction of ferromagnetic-metallic phase undergoes glass-like kinetic arrest and coexists at low temperature with the equilibrium antiferromagnetic-insulating phase. We show here through resistivity measurements that the residual arrested ferromagnetic-metallic fraction can be converted to the equilibrium antiferromagnetic-insulating phase by successive annealing at higher temperatures, possibly through heterogeneous nucleation of equilibrium phase. Significantly, larger fractions of this glassy ferromagnetic-metallic phase can be obtained by cooling in higher fields and larger conversion to equilibrium antiferromagnetic-insulating phase results.

Glasses form below a temperature where dynamics is arrested, preserving the high-temperature structure while avoiding the first-order liquid-solid transformation at the higher temperature $T_C$. Rapid cool-down is needed to arrest the transformation kinetics for metallic alloys, while glass-formers like O-terphenyl vitrify easily because the kinetics is arrested at a temperature $T_K$ that is closer to $T_C$[1-3]. Arrest of kinetics can also inhibit a first-order process where both the phases on either side of the transition have long-range structural, including magnetic, order. It has been shown recently that glass-like arrest of kinetics intervenes in first-order magnetic transformations and results in coexisting phases with competing magnetic orders[4-8]. In $La_{0.5}Ca_{0.5}MnO_3$ manganite, the high-temperature long-range ferromagnetic-metallic phase transforms to antiferromagnetic-insulating phase, although a fraction of ferromagnetic-metallic phase coexists at low temperature[9]. Magnetization measurements show that this ferromagnetic-metallic phase has undergone glass-like kinetic arrest and coexists at low temperature with the equilibrium antiferromagnetic-insulating phase, similar to other half-doped manganites[8,10]. We show here through resistivity measurements that the residual arrested ferromagnetic-metallic fraction can be converted to the equilibrium antiferromagnetic-insulating phase by successive annealing at higher temperatures[1]. The restoration of kinetics in the arrested phase presumably results in heterogeneous nucleation of equilibrium phase. We obtained larger fractions of this glassy ferromagnetic-metallic phase by cooling in higher fields and found better conversion to equilibrium antiferromagnetic-insulating phase, reminiscent of devitrification of structural glasses into crystallites. The $T_C$ can be varied over a wide range with magnetic field and the difference between $T_K$ and $T_C$ can be tuned in the same sample. This advantage may help shed light on the physics of structural glasses.

Although glasses have been used for many centuries and variety of materials including monatomic metallic liquids are capable of glass formation[11], a quantitative understanding of the glass transition remains a major scientific challenge. Glasses form by freezing the structure of a liquid; but all amorphous solids are not glass[1,2]. The general principle underlying diverse glasses is the arrest of the kinetics of the first-order transition, which connects two symmetrically incompatible orders, freezing the higher energy state into a non-ergodic state at low temperature. The concept of glass-like arrest of kinetics has recently been invoked to successfully explain various anomalies in disorder broadened first-order magnetic transitions as a function of field and temperature for materials like colossal magnetoresistance manganites[5-8,10]. These have long-range magnetic orders on either side of the transition. However, the high-temperature phase persists in the low-temperature region where it is energetically unstable, and the lack of dynamics triumphs over thermodynamics. In pure systems, the first-order transition can occur along a sharply defined $(H_C, T_C)$ line in the 2-control variable (H, T) space. Due to disorders, different regions having length-scale of the order of the correlation-length can have different $T_C$ resulting in the transition line broadening into a band; as also do the lines corresponding to supercooling (H*, T*) and superheating (H**, T**) limits[12,13]. Similarly, if the glass-like kinetic arrest is to occur below a $(H_K, T_K)$ line for a given cooling rate in the pure system, disorder broadens this into a band consisting of quasi-continuum of $(H_K, T_K)$ lines[5,7,8,10,14,15]. By traversing the 2-control variable H-T space, tunable coexisting fractions of arrested and equilibrium phases have been observed because of this disorder broadening. Devitrification is an evidence of glassy state[1-3] which is recently demonstrated for such glassy magnetic state of variety of systems[7,8,10,14]. The present study focuses on a half-doped manganite, $La_{0.5}Ca_{0.5}MnO_3$, similar to the sample where Loudon et al. have proposed a 'nucleation and growth' process for transformation from high-temperature ferromagnetic-metallic (FMM) to low-temperature antiferromagnetic-insulating (AFI) phase[9]. Coexistence of FMM with AFI regions at low temperature was shown, and here we explore their origin and nature.

Polycrystalline $La_{0.5}Ca_{0.5}MnO_3$ sample has been prepared through a well-established chemical route known as 'pyrophoric method'. High purity (>99.9%) $La_2O_3$, $CaCO_3$ and $C_4H_6MnO_4.4H_2O$ are taken in stoichiometric quantities as starting materials. These materials are dissolved in aqueous nitric acid and the resulting solutions are mixed together with triethanolamine (TEA). The complex solution is heated to dehydrate and decompose leaving behind organic-based, black, fluffy precursor powder. This dried mass is then grounded to fine powder, palletized and then calcined at $1000^0C$ for 3 hrs in oxygen atmosphere. The powder x-Ray diffraction (XRD) was carried out using an 18 kW Rigaku Rotaflex RTC 300 RC diffractometer with Cu-$K_\alpha$ radiation. Rietveld profile refinement of XRD pattern confirms that the sample is in single phase without any detectable impurity and crystallizes in orthorhombic structure with 'pnma' space group. The resistivity and magnetic measurements are performed using commercial set-ups (14Tesla-PPMS-VSM, M/s. Quantum Design, USA).

Figure-1(a) shows the magnetization as a function of temperature measured in 1 Tesla field under various protocols. The hysteresis between field-cooled cooling (FCC) and field-cooled warming (FCW) paths indicates a disorder-broadened first-order transition from FMM to AFI with reducing temperature. However, a substantial magnetization in the low temperature antiferromagnetic phase, similar to Ref. 9, suggests the persistence of ferromagnetic phase. Because of disorder-broadening of (H*, T*) and (H$_K$, T$_K$) lines into bands, the fraction of glass-like arrested state can be controlled by cooling in different fields[5-8, 15]. Based on the measurements similar to those reported earlier[7,8] in the glass-like kinetically arrested magnetic systems, we propose that the AFI state is in equilibrium at low temperature and the FMM phase fraction exists as kinetically arrested glassy or non-ergodic state. We can collect larger fractions of glass-like arrested FMM phase at 5K by cooling in 6T field and then reducing field to 1T. This glass-like arrested FMM state

devitrifies on heating as depicted in Fig.1 by the rapid fall in magnetization, of the 6T-cooled states, which approach the equilibrium AFI phase while warming. This half-doped manganite should have the spin-aligned value of 3.5 $\mu_B$/Mn accompanied by metallic conductivity[9]. The observed 1T-FCC magnetization of 0.61 $\mu_B$/Mn at 5K can be attributed to a frozen FMM phase fraction of about 17%, which is close to the percolation threshold for electrical conductivity. Hence, around this FMM phase fraction the drastic resistivity changes are a more sensitive tool, compared to the magnetization, to probe small changes in the phase fractions. Figure 1(b) shows the heating and cooling cycles of the zero-field resistivity. Apart from the expected thermal hysteresis, the decrease in resistivity with the decrease in temperature below ~ 70K reflects the presence of FMM phase fraction in AFI matrix. Similar to magnetization, the zero-field resistivity also shows larger fraction of FMM phase when the sample is cooled in 6T (Fig. 1b). Different values of magnetization and resistivity in the same measurement temperature (5K) and field (1T for magnetization and zero for resistivity) indicate the presence of non-ergodic states. The concomitant sharp decrease in magnetization and increase in resistivity, of the 6T cooled state, around 20 K as shown in Fig. 1 (a) and (b) results when this arrested non-ergodic FMM phase devitrifies to AFI phase on warming[8].

This glass-like arrested FMM state resembles the conventional structural glasses in various respects. We now show that it partially crystallizes to equilibrium AFI state after annealing, similar to the route followed to produce nano-crystals within the metallic glasses or to the formation of glass ceramic[1]. Fig 2 (a) shows resistivity in zero-field while warming after cooling from 320K in zero field and again after a temperature cycle from 5K to 140K to 5K without applying any field. The increase in resistivity after annealing to 140K indicates that the system is ripening to its equilibrium state similar to crystallization, in this case to AFI phase at low temperature. For conventional systems such an effect can take place from heterogeneous nucleation process. When contrasted with the initial cool-down, we now have an abundance of AFI seeds to trigger this process when energy becomes available to the system[16]. It is well known that larger fraction of a glass can be devitrified from a partially crystalline state by successive annealing[1]. Similar effect of successive annealing at progressively higher temperatures has been observed in resistivity (as also in magnetization, not shown here). Fig. 2(b) shows the resistivity in zero-field at 5K as a function of progressively higher annealing temperature after cooling the sample to 5K in zero field. Each temperature cycle to progressively higher temperature produces larger factions of AFI phase and the system approaches the equilibrium state overcoming a hierarchy of barriers. Back conversion to FMM phase occurs, however, as the annealing temperature is raised above 150K.

We now investigate whether the devitrification and heterogeneous nucleation is influenced by the initial glass-like kinetically arrested FMM phase fraction. This fraction can be tuned by cooling in different fields[8]; is about 17% when cooled in a field of 1T, but rises to about 90% when cooled in 6T [Figure 1(a)]. This fraction remains fixed when the field is reduced to zero at 5K. We show in Figures 3(a) and 3(b) the results of annealing to 140K on these two cases. The initial higher FMM phase fraction for higher cooling field is evidenced by lower resistivity at 5K in zero field. However, after annealing to 140K the behaviour reverses, the lower resistivity of the 6T-cooled state becomes much higher compared to 1T-cooled state and exceeds the measurement range of the instrument below 38K. Similarly, the 1T-cooled state, initially having lower resistivity compared to zero field cooled state, shows higher resistivity after annealing to 140K [Figures 2(a) and 3(a)]. The effect of successive annealing on 6T- and 1T-cooled states is shown in Figure 3(c). Annealing to temperatures higher than 80K clearly brings out a conversion to larger equilibrium AFI state when the starting glass-like kinetically arrested FMM phase fraction is larger. The effect is drastic for annealing temperature of 120K and beyond. This result may appear counterintuitive but may have some connection with the folklore that "hot water can freeze faster than cold"[17] or the everyday experience that bigger pieces of glass are more susceptible to shattering when dropped.

The ease with which field (and temperature) can be varied contrasts with the current experimental efforts involved in varying pressure (and temperature) to explore the energy landscape of glassy systems[11,18]. Since $T_C$ varies strongly with magnetic field, so does the difference between $T_K$ and $T_C$; cooling in different fields allows us to arrest varying fractions of glass-like FMM phase without changing the rate of cooling[5-8]. Such glass-like kinetic arrest of first-order magnetic transitions, especially where the transition is broadened by disorder, should enable studying the physics of structural glasses.

**Acknowledgement:** DST, Government of India is acknowledged for funding the 14 Tesla-PPMS-VSM.

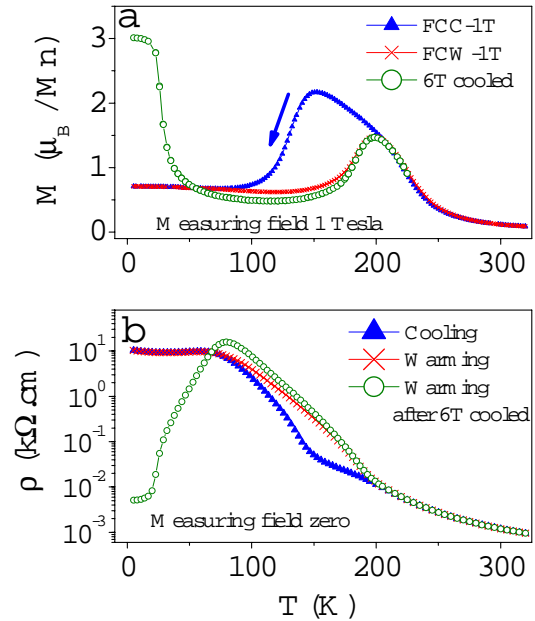

**Figure 1:** Temperature (T) dependence of Magnetization (M) in 1 Tesla field and resistivity (R) in zero field of La$_{0.5}$Ca$_{0.5}$MnO$_3$ measured under different protocols. **a,** M vs. T while cooling (FCC) in 1T field from 320K to 5K and again while warming (FCW) from 5K shows the thermal hysteresis accompanying the first-order ferromagnetic to antiferromagnetic transition. After cooling the sample in 6T field the field is reduced isothermally to 1T at 5K and M is measured while warming. The large value of M at 5K reflects a dominant arrested ferromagnetic phase, and its devitrification starts around 20K. **b,** R vs. T while cooling in zero field from 320K to 5K and again while warming from 5K shows the thermal hysteresis accompanying the first-order metallic to insulating transition. After cooling the sample in 6T field, the field is reduced isothermally to zero at 5K and R is measured while warming. The low value of R at 5K reflects a large fraction of arrested metallic phase, and its devitrification starts around 20K. Thus the high temperature ferromagnetic-metallic phase (FMM) undergoes a first-order transition to antiferromagnetic-insulating (AFI) phase at low temperature. However, multivalued M at 5K and 1T and also different values of R at 5K in zero field indicate the presence of non-ergodic states. Their devtrification while warming is indicated by sharp changes in M and R around 20K.

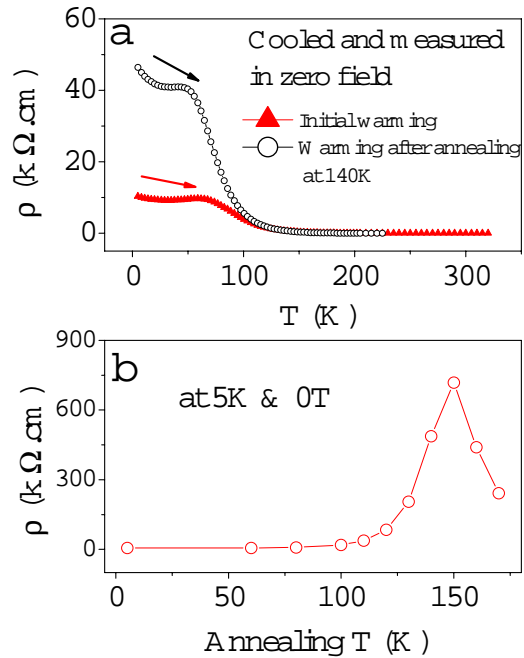

**Figure 2:** Resistivity as a function of temperature in zero field. **a,** Resistivity while warming from 5K after cooling from 320K in zero field. Again the sample is cooled from 320K to 5K, then warmed to 140K and cooled back to 5K. Resistivity is measured while warming after this temperature cycle. The large difference in the resistivity values at low temperature, between the initially cooled state and that after annealing to 140K, arises from the devitrification through the heterogeneous nucleation process. During initial cool-down the entire sample was FMM and the AFI phase formed through homogeneous nucleation. After warming from 5K and annealing at 140K, seeds of AFI phase enabled heterogeneous nucleation, reducing the residual FMM fraction. **b,** Resistivity values at 5K as a function of annealing temperatures. After cooling the sample to 5K, resistivity is measured while cycling the temperature between 5K and the successive higher annealing temperatures. Progressively higher resistivity values at 5K after successive annealing indicate additional devitrification toward equilibrium AFI phase. This successive annealing produces higher resistivity at 5K than the single annealing shown in **a**. Annealing above 150K causes back conversion to FMM phase and resistivity drops. The cooling/heating rates and the temperature interval between the data points are maintained all through these measurements.

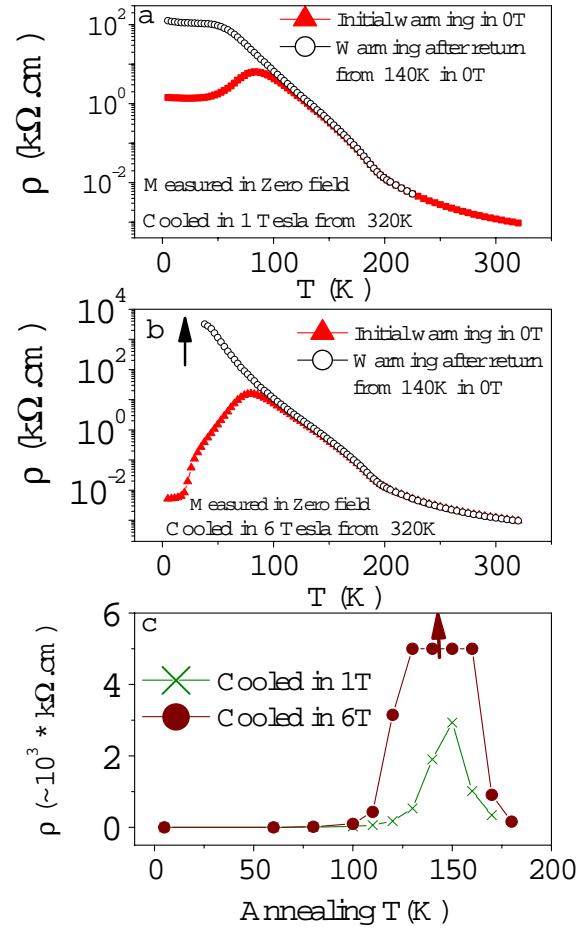

**Figure 3:** Effect of larger initial fractions of the glass-like FMM phase produced after initial cooling in 1T and 6T on the single-shot annealing, and on successive annealing, is studied through zero-field resistivity measurements. **a,** We cool from 320K in 1T to 5K and then isothermally reduce the field to zero. Resistivity in zero field is measured while warming from 5K. The sample is again cooled from 320K to 5K in 1T and then after isothermally reducing the field to zero, warmed to 140K and cooled back to 5K. Resistivity is measured while warming after this temperature cycle without changing the field condition (maintained as zero). **b,** Resistivity in zero field while warming similar to the protocol of **a** but this time the cooling field 6T is used during cooling from 320K to 5K. The lower value of R at 5K after cooling from 320K in 6T and then isothermally reducing the field to zero, reflects a larger fraction of arrested metallic phase. The large difference in the resistivity values between the initially cooled state and that after annealing to 140K in both these cases, as in figure 2a and 2b, arises from the heterogeneous nucleation process. The resistivity below 38K exceeds the measurement range of the instrument. **c,** Resistivity values at 5K as a function of annealing temperatures. After cooling the sample to 5K once in 1T (and whole process is repeated with 6T), resistivity is measured while cycling the temperature between 5K and the successively higher annealing temperatures. Progressively higher resistivity values at 5K after successive annealing indicate additional devitrification toward equilibrium AFI phase. This successive annealing produces higher resistivity at 5K than the single annealing of the respective cases, and the 6T-cooled case gives the largest AFI fraction. Its resistivity at 5K after annealing to 130K, 140K, 150K and 160K exceeds the measurement range of the instrument. Annealing above 150K causes back conversion to FMM phase and resistivity drops.